\documentclass[acmsmall,screen,nonacm]{acmart}

\usepackage[utf8]{inputenc}
\usepackage{graphicx}
\usepackage{subcaption}

\makeatletter
\renewcommand{\keywordsname}{%
    \textbf{\textit{Keywords --}}\@gobble}
\makeatother

\begin{document}

\title[TSAE: Quantifying Communication-Based Synchrony in Collaborative VR]{Time-Dynamic Semantic-Activity Entanglement: Quantifying Communication-Based Synchrony in Collaborative VR}

\author{Pooja Pol}

\affiliation{%
    \institution{Technische Hochschule Augsburg,}
    \institution{TTZ Landsberg am Lech}
    \city{Augsburg}
    \country{Germany}
}

\email{pooja.pol@tha.de}

\author{Jianing Zhang}
\affiliation{%
    \institution{Technische Hochschule Augsburg,}
    \institution{TTZ Landsberg am Lech}
    \city{Augsburg}
    \country{Germany}
}
\email{jianing.zhang@tha.de}

\author{Qing Huang}

\affiliation{%
    \institution{Technische Hochschule Augsburg,}
    \institution{TTZ Landsberg am Lech}
    \city{Augsburg}
    \country{Germany}
}
\email{qing.huang@tha.de}

\begin{abstract}
\noindent\textbf{\textit{Abstract --}} Collaborative virtual reality (VR) produces rich multimodal traces of team behavior, yet practical methods for characterizing time-varying team communication remain limited. We present time-dynamic semantic-activity entanglement (TSAE), inspired by the entanglement concept introduced by \citeauthor{gloor2022entanglement}~\cite{gloor2022entanglement}, to describe communication-based team synchronization within discrete time windows. TSAE is modeled by combining relative speaking activity with contextual semantic alignment for participant pairs, which is aggregated into participant-level trajectories and summarized through the Gini coefficient to understand the distribution of team-level communication synchrony. This approach is employed with German-language transcripts from fifteen collaborative VR gameplay sessions. Transcript utterances are represented using a long-context embedding model with a late-chunking strategy and analyzed within fixed game-time windows. A representative session illustrates shifting dyadic alignment and low-to-moderate participant contributions, with sporadic periods of high entanglement. Across fifteen game sessions, the metrics indicate complementary insights, offering a qualitative contrast to the core–periphery pattern reported in the foundational work. Hence, these descriptive findings support the study of communication as a dynamic process and provide a mathematically grounded approach to analyze its evolving distribution in shared interactive VR spaces.
\end{abstract}


\maketitle

\section{Introduction and Background}

Collaborative virtual reality (VR) requires participants to coordinate
actions and maintain a shared understanding of
the task and of each other
\cite{ghamandi2024unlocking,paulsen2024collaborativevr}.
Team processes in collaborative problem solving (CPS) tasks are non-static, so aggregate end-state summaries can miss relevant temporal dynamics. Characterizing changes in individual and interpersonal behavior in CPS can be related to different phases of team collaboration \cite{delice2019advancing,lehmann2024multimodal}. Additionally, studies of workspace layouts and avatar representations show the influence of interactive space designs on collaborative experiences \cite{enriquez2024layout,fink2024avatars}. Research on social interaction in physical and virtual settings examines how spatial behavior, audio, and available social cues shape interaction~\cite{williamson2022digital,lechappe2025understanding}. Among multiple modalities, such as gaze, or movement, explored in these studies on social synchronization, speech offers an accessible starting point for identifying episodes of group alignment. This motivates time-resolved measures of team-level interaction while preserving information on individual and dyadic contributions.

Our methodological predecessor is ~\citeauthor{gloor2022entanglement}'s entanglement metric
\cite{gloor2022entanglement}, which investigates team synchronization of communication network time series, primarily grounded in communication volume. Their organizational studies associated teams with higher performance having a strongly entangled communication core with few members and less entangled periphery members. But their email networks are directional, whereas in free-flowing VR, conversational addressees can be ambiguous. Drawing on research on team flow and the notion of being \texttt{in sync}~\cite{van2018conceptualization}, we adapt entanglement for undirected dyads in a VR setup.

We introduce TSAE, time-dynamic semantic-activity entanglement, which combines relative speaking activity and the semantic context of what is spoken between team members as a time series. Dyadic scores are first aggregated into participant-level trajectories, and the Gini coefficient summarizes how entanglement is distributed in the team. An empirical evaluation of a collaborative VR dataset is carried out and enriched with an interpretation of these metrics. Together, the formulation and analysis provide a descriptive and mathematically grounded method to examine the evolving structure of communication in shared interactive spaces.

\begin{figure*}[t]
    \centering
    \includegraphics[width=\textwidth]{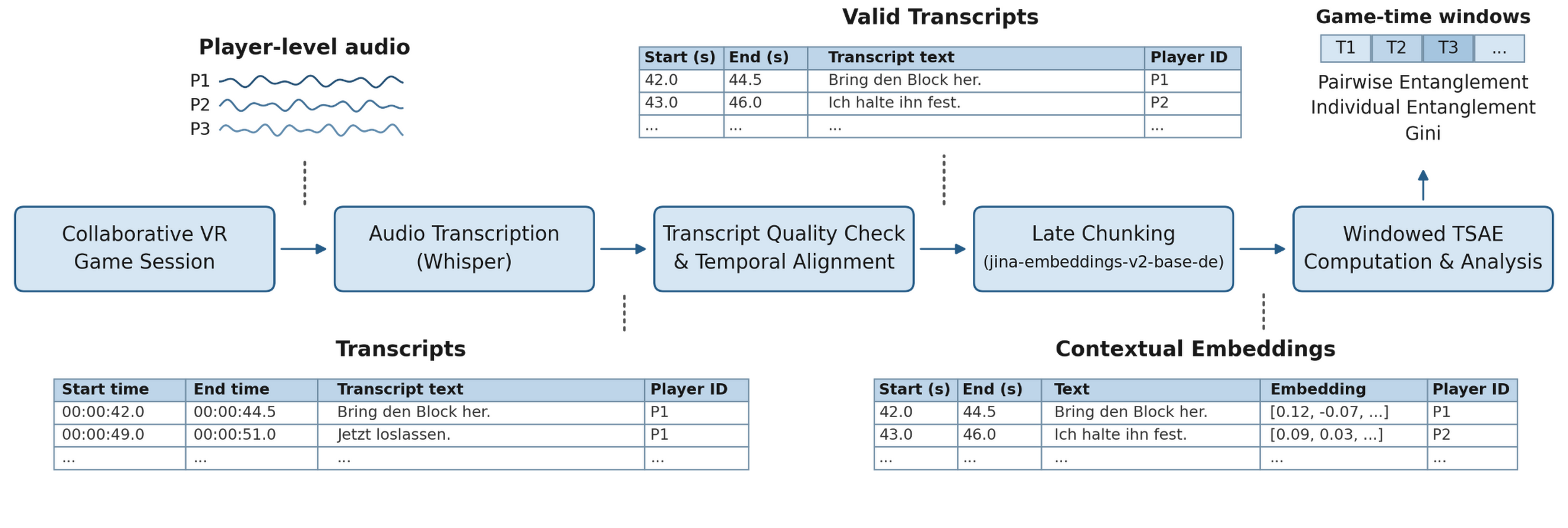}
    \caption{Pipeline computing entanglement (TSAE) and Gini metrics.}
    \Description{Five blocks, showing the flow of player-level audio signals and their step-by-step processing into the final computation of the entanglement and Gini metrics. The intermediate results obtained after each block are displayed over the arrow, showing some synthetic examples of the processing steps.}
    \label{fig:tsae_pipeline}
\end{figure*}

\section{Method}
\label{sec:methodology}

\subsection{Time-Dynamic Semantic-Activity Entanglement: Formulation}
\label{sec:tsae_formulation}

Let \(P=\{p_1,\ldots,p_n\}\) be the players in a team. Let the game timeline \(T=\{T_1,\ldots,T_m\}\) be divided into \(m\) discrete time windows. For a participant \(p_i\) within a time window \(T_k\), let the count of spoken words be \(a_i^k\), the corresponding set of utterances be \(U_i^k\), and \(\mathbf{v}(u)\in\mathbb{R}^d\) represent the late-chunked embedding of an utterance \(u\). We can then represent the participant's average window-level semantics by $\bar{z}_i^k = \frac{1}{|U_i^k|} \sum_{u\in U_i^k} \mathbf{v}(u)$.
We use mean pooling as it provides a stable semantic representation of a player while avoiding the biases induced by extrema-based pooling.

For a participant pair \((p_i,p_j)\) in time window \(T_k\), we define three components:
\begin{equation}
A_{ij}^k = \frac{a_i^k+a_j^k}{\sum_{\ell=1}^{n}a_\ell^k}, \quad
C_{ij}^k = \frac{\bar{\mathbf{z}}_i^k\cdot\bar{\mathbf{z}}_j^k}{\|\bar{\mathbf{z}}_i^k\|_2 \|\bar{\mathbf{z}}_j^k\|_2}, \quad
\Delta_{ij}^k = \|\bar{\mathbf{z}}_i^k-\bar{\mathbf{z}}_j^k\|_2\,\text{,}
\label{eq:tsae_components}
\end{equation}
where \(A_{ij}^k\) is the \(p_i\)-\(p_j\) pair's share of the total speaking activity of a team, \(C_{ij}^k\) is their cosine similarity, and \(\Delta_{ij}^k\) penalizes their geometric separation. \(C_{ij}^k\) captures angular semantic alignment measured by values in the range \([-1,1]\), where the upper bound \(1\) indicates identical direction in the embedding space. \(\Delta_{ij}^k\) penalizes the absolute separation between the non-normalized participant centroids, including differences not represented by angular similarity alone. 

Pairwise entanglement (TSAE) is then given by
\begin{equation}
E_{ij}^k=\frac{A_{ij}^k\,C_{ij}^k}{1+\Delta_{ij}^k}\,\text{.}
\label{eq:pairwise_tsae}
\end{equation}
A large positive pairwise TSAE requires both substantial joint activity and contextual semantic alignment.
Adding one to the denominator prevents division by zero and converts Euclidean distance into an attenuation factor that progressively reduces TSAE as the participant centroids become more separated.

Individual participant-level entanglement is obtained by averaging over the pairwise values:
\begin{equation}
E_i^k=\frac{1}{n-1}\sum_{j\neq i}E_{ij}^k\,\text{.}
\label{eq:individual_tsae}
\end{equation}
Using the fixed denominator of \(n-1\) possible teammates ensures a consistent scale across all window sizes.

When all \(E_i^k\) are finite, non-negative, and their sum is positive, we can compute their normalized individual shares as 
\(\pi_i^k=E_i^k/\sum_{\ell}E_\ell^k\), such that
\(\sum_{i=1}^{n}\pi_i^k=1\). 
We can then characterize communication synchrony at the team level as a distribution of these shares among participants, defined as the Gini coefficient
\begin{equation}
G_k=\frac{1}{2n}\sum_{i=1}^{n}\sum_{j=1}^{n}|\pi_i^k-\pi_j^k|\,\text{.}
\label{eq:tsae_gini}
\end{equation}
Lower values indicate a more even distribution of entanglement and higher values indicate more inequality and show concentration among fewer participants.

\subsection{VR Task and Dataset}

Teams of five to eight adults in a co-located VR environment complete a CPS task. Players coordinate to achieve a common game objective, which naturally elicits verbal team communication. Our dataset comprises fifteen valid German-language sessions (one team each) with complete game timestamps and player-level \(48\,\mathrm{kHz}\) mono audio recorded via Meta Quest Pro head-mounted displays (HMDs)~\cite{metaquestpro}. Each game-play session averages about \(25\) minutes. All data is anonymized with informed consent and privacy constraints preclude demographic or prior player-relationship metadata.

\subsection{Data Processing and Pipeline}

Figure~\ref{fig:tsae_pipeline} outlines the pipeline for computing the metrics proposed in this study, illustrating synthetic examples of transcripts and embeddings. Player audio channels are transcribed using Whisper~\cite{radford2022whisper}. The timestamped German utterances are manually verified for completeness and aligned to a common game clock to retain concurrent speech. We encode utterances using the Jina v2 model~\cite{jinabaseDE2024} with late chunking~\cite{mohr2024bilingual,gunther2024late}. This approach can be more meaningful in collaborative dialogue because utterances are short local units whose meaning often depends on surrounding conversational context that could be lost while encoding shorter chunks independently. Finally, the late-chunked embeddings and participant word counts are used to compute the aforementioned metrics in valid time windows. Silence or inactivity is considered as zero activity volume, thus yielding zero computed entanglement.

\begin{figure*}[t]
    \centering
    \includegraphics[width=\textwidth]{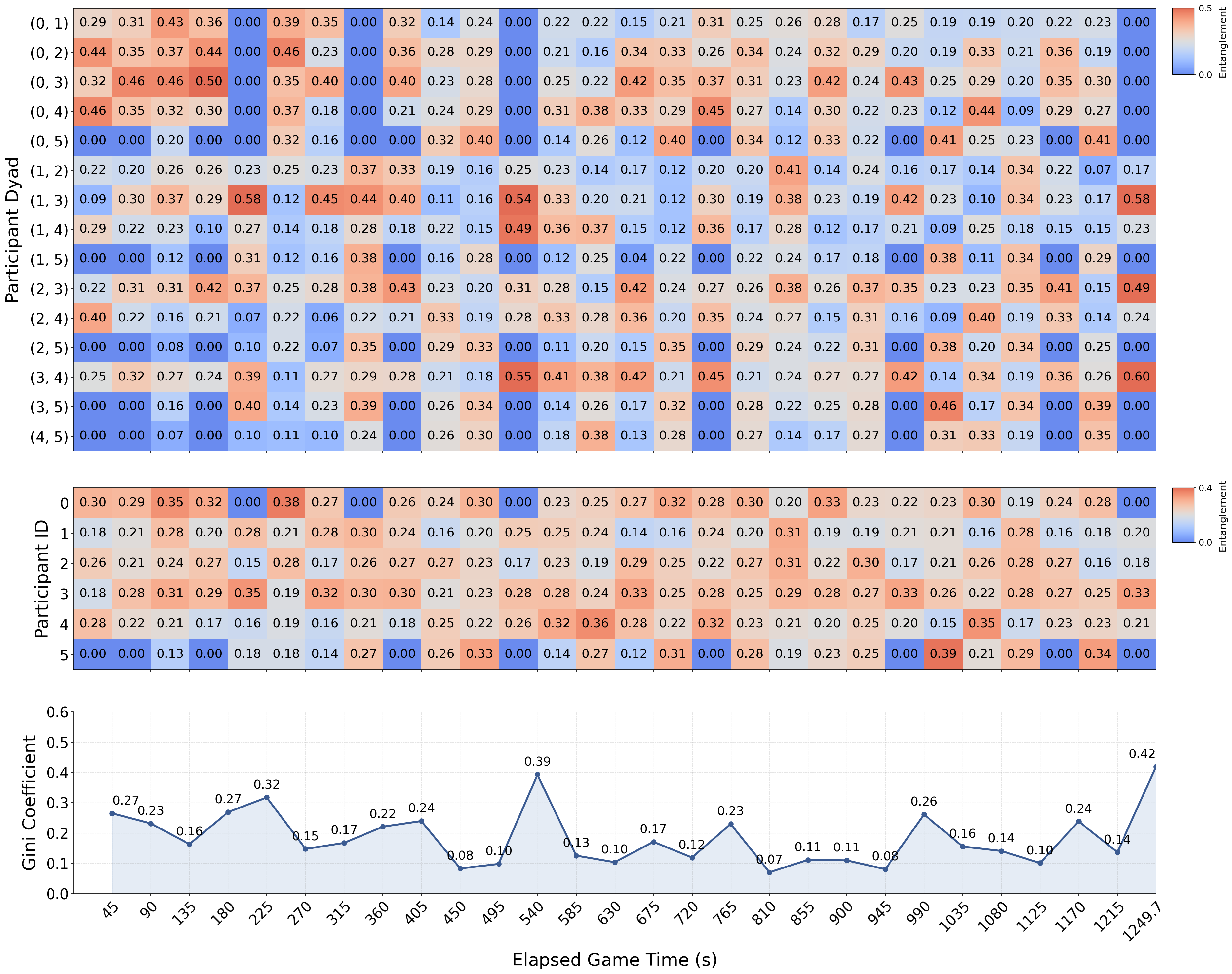}

    \caption{Pairwise entanglement (top), participant-level individual entanglement (middle),
    and the distribution of individual entanglement, measured as the Gini coefficient (bottom) in a
    representative session over \(45\,\mathrm{s}\) game-time windows.}
    \label{fig:tsae-combined}

    \Description{Three vertically aligned panels share an elapsed game-time axis with 45-second windows. The top heatmap shows pairwise TSAE for all participant pairs in this session; the middle heatmap shows individual TSAE for six participants. Cell annotations report entanglement values. The bottom line plot shows the Gini coefficient derived from individual entanglement distribution over time.
    }
\end{figure*}

\section{Results}

\subsection{Communication-based Entanglement and Gini Coefficients in a Representative Session}
\label{sec:comm_sync_res}

Figure~\ref{fig:tsae-combined} shows time-resolved TSAE and Gini metrics computed for one specific six-participant session. The heatmaps report entanglement values, with warmer colors indicating high entanglement and cooler colors indicating low entanglement. Dyadic entanglement (top) varies across
windows, with elevations for dyad \((1,3)\) at \(225\,\mathrm{s}\) (\(E_{ij}=0.58\)) and dyad \((3,4)\) at
\(540\,\mathrm{s}\) (\(E_{ij}=0.55\)). Participant-level scores (middle) also show similar patterns, with participants~0 and~4 relatively elevated at \(1080\,\mathrm{s}\). Most Gini coefficients (bottom) lie between \(0.07\) and \(0.27\), indicating relatively even participant entanglement. Higher concentration occurs at \(540\,\mathrm{s}\) (\(G=0.39\)) and in the final window (\(G=0.42\)). The heat in the entanglement values is observed to change intermittently, resulting in rapid swapping of alignment over time, rather than a consistent pattern of communication synchrony. The distributional team-level Gini timeline reveals convergent observations, suggesting a distributed communication structure, with some peaks showing concentration among fewer participants. Hence, the three views jointly explain the evolving structure of team communication.

\begin{figure*}[t]
    \centering
    \Description{Cross session analysis visualization. The first sub-figure shows the sensitivity of the session's median Gini coefficient values at different time windows. The second sub-figure shows a statistical representation of the entanglement shares, ordered from highest to lowest ranks of players, and their corresponding contributions through descending Lorenz curves at a 45 second temporal resolution.}
    \begin{subfigure}[t]{0.49\textwidth}
        \centering
        \includegraphics[width=\linewidth]{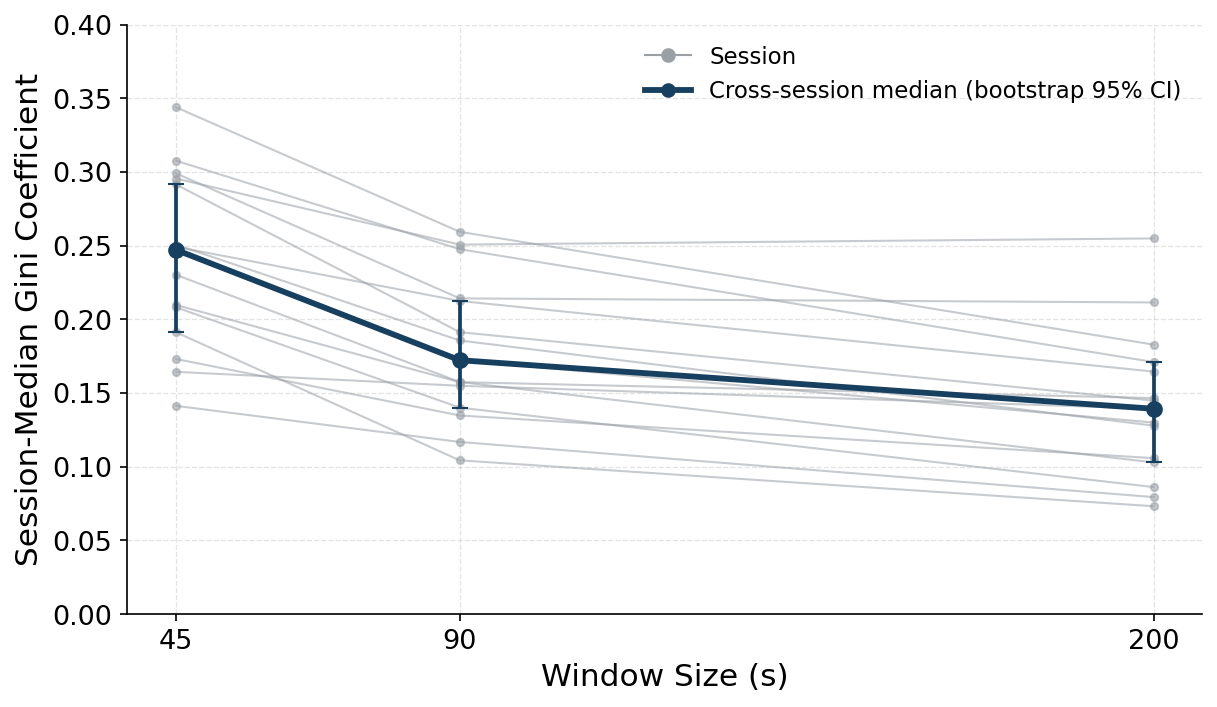}
        \caption{Temporal-resolution sensitivity of session-level median Gini coefficients at \(45\,\mathrm{s}\), \(90\,\mathrm{s}\), and \(200\,\mathrm{s}\).}
        \label{fig:median_gini}
    \end{subfigure}
    \hfill
    \begin{subfigure}[t]{0.49\textwidth}
        \centering
        \includegraphics[width=\linewidth]{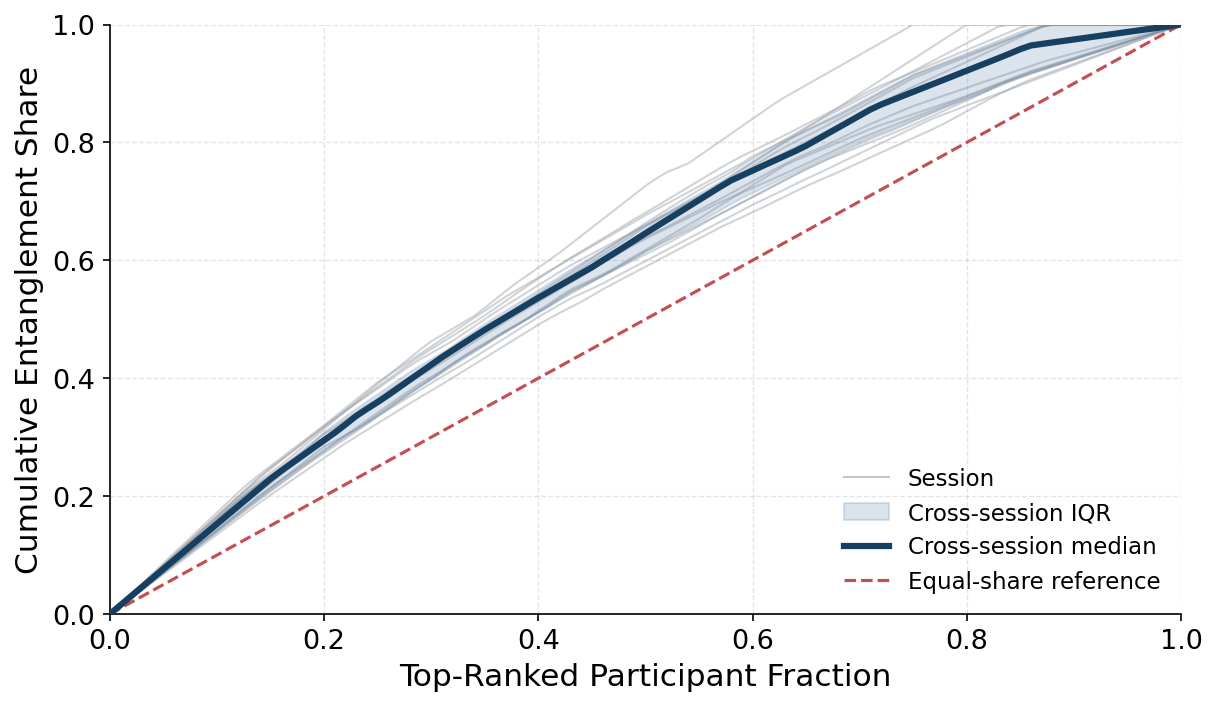}
        \caption{Descending Lorenz curves of cumulative entanglement shares at the \(45\,\mathrm{s}\) resolution.}
        \label{fig:topk_share}
    \end{subfigure}
    \caption{Cross-session patterns of Gini coefficients and cumulative entanglement shares.}
    \label{fig:cross_sess}
\end{figure*}

\begin{table*}[t]
    \centering
    \caption{Session-level Gini coefficients computed from the
    Lorenz curves in Figure~\ref{fig:topk_share}.}
    \label{tab:session_curve_gini}

    \small
    \setlength{\tabcolsep}{1.5pt}
    \renewcommand{\arraystretch}{1.2}

    \begin{tabular*}{\textwidth}
        {@{\extracolsep{\fill}}l*{15}{c}@{}}
        \toprule

        \textbf{Session}
        & 1 & 2 & 3 & 4 & 5
        & 6 & 7 & 8 & 9 & 10
        & 11 & 12 & 13 & 14 & 15 \\

        \textbf{Participants (\(n\))}
        & 7 & 6 & 6 & 7 & 8
        & 8 & 7 & 8 & 8 & 7
        & 7 & 5 & 5 & 8 & 7 \\

        \textbf{Gini coefficient}
        & 0.179 & 0.288 & 0.181 & 0.211 & 0.296
        & 0.255 & 0.307 & 0.217 & 0.355 & 0.139
        & 0.268 & 0.172 & 0.303 & 0.259 & 0.192 \\

        \bottomrule
    \end{tabular*}
\end{table*}

\subsection{Cross-Session Analysis}
\label{sec:cross_sess_res}

We analyze the distributional patterns in fifteen sessions using the finite-team correction of the Gini coefficient in Equation~\ref{eq:tsae_gini} as \(G_k^{\ast}=\frac{n}{n-1}G_k\). This ensures a common \([0,1]\) normalization across teams of varying sizes. In Figure~\ref{fig:median_gini},
gray lines show each session's corrected Gini coefficient calculated at varying time resolution, and the blue line and error bars show the cross-session median and \(95\%\) bootstrap confidence intervals. The cross-session median decreases from approximately \(0.25\) at \(45\,\mathrm{s}\) to \(0.17\) at \(90\,\mathrm{s}\) and \(0.14\) at \(200\,\mathrm{s}\). Most sessions follow this trend, where longer windows smooth transient differences in participant entanglement and reduce apparent concentration. 

Figure~\ref{fig:topk_share} summarizes individual entanglement shares using descending Lorenz curves, with participants ranked from highest to lowest by their accumulated shares. Here, gray lines show the session-level Lorenz curves and the blue line aggregates them into a cross-session median curve. The diagonal denotes equal shares, and the elevated curves indicate the extent of higher concentration. The blue curve shows that the top-ranked \(20\%\), \(40\%\), and \(60\%\) of participants account for approximately \(31\%\), \(54\%\), and \(74\%\) of total entanglement, respectively. Thus, the top \(40\%\) hold just over half the total, showing some inequality but relatively distributed entanglement. This is supported by the Gini coefficients computed from the area under the individual-session Lorenz curves reported in Table~\ref{tab:session_curve_gini}. They range from \(0.139\) to \(0.355\), indicating a pattern of uniformly distributed entanglement and some variation in concentration throughout the fifteen sessions.

\section{Implications and Discussion}

The findings of our study support examining team communication as a temporally varying distribution of activity and semantic alignment, connecting individual and dyadic behavior to the team-level perspective motivated in prior work~\cite{delice2019advancing,lehmann2024multimodal}. The representative session shows shifting dyadic and participant entanglement with intermittent concentration, which
could reflect task-specific contributions and warrants examination against the ongoing activity. The single session results are consistent with the implications of the cross-session patterns indicating broadly distributed shares. This becomes tractable through the study of the Lorenz curves, the corresponding Gini coefficients and the heatmaps. Our study, thus, offers a descriptive contrast to \citeauthor{gloor2022entanglement}, who associated higher performance with greater entanglement dispersion, interpreted as a core-periphery pattern.

Several processing choices can be explored to allow for more granular interpretations. Because our analysis uses manually quality-checked transcripts, robustness to uncorrected automatic speech recognition (ASR) errors remains untested. Generalization may be limited by factors such as the German-language dataset and limited availability of demographic information. 

By combining speaking activity with contextual semantic alignment, TSAE provides a time-dynamic speech-based analytical layer to examine team communication, without establishing subjective team flow or
collaboration quality. Examining the activity and semantic components of communication can clarify whether the observed changes reflect shifts in verbal participation, semantic alignment, or both. Comparisons with other communication or behavioral measures could investigate how these entanglement patterns are linked to other aspects of collaborative activity, helping assess the contribution of contextual information in descriptions of social interaction. Across different CPS tasks, the examination of TSAE trajectories alongside participant reports, behavioral annotations, or task outcomes could provide a basis for testing hypotheses and thus enrich our understanding of team dynamics and effective coordination. 

\section{Conclusion}
We present time-dynamic semantic-activity entanglement (TSAE) as a contextual metric to quantify communication-based synchronization. The idea involves combining speaking activity with semantic similarity from transcript data. The entanglement metric provides explainable views of speaking activity and communication synchrony in a team at different levels of granularity such as dyadic, participant-level and distributional. Furthermore, the Gini coefficient is another derived metric that provides a team-level characterization of communication synchrony. Contrary to the realization of~\citeauthor{gloor2022entanglement}~\cite{gloor2022entanglement} about team communication, our analysis shows relatively uniformly distributed entanglement with intermittent redistribution of high entanglement over time. Cross-session analysis shows low-to-moderate inequality in entanglement shares, consistent with the distributional observations of the representative session, and Gini coefficients at variable temporal resolutions further highlight the influence of the observation timescale. Although validation against explicit hypotheses is needed to establish what these descriptive patterns signify in CPS, this work demonstrates an interpretable approach to investigate time-resolved communication in groups.

\begin{acks}
This research was supported by Technische Hoch\-schule Augsburg and the Hightech Agenda Bavaria, funded by the Free State of Bavaria, Germany. We thank Lumium GmbH for VR game development, hardware and experimental setup, game session organization and execution, and data access. We also thank our colleagues at the Data Science und Autonome Systeme Technologietransferzentrum (TTZ) Landsberg for their insightful discussions throughout this research.
\end{acks}

\bibliographystyle{ACM-Reference-Format}
\bibliography{references}

@inproceedings{ghamandi2024unlocking,
  author        = {Ghamandi, Ryan and Kattoju, Ravi Kiran and Hmaiti, Yahya and Maslych, Mykola and Taranta, II, Eugene and McMahan, Ryan P. and LaViola, Jr., Joseph J.},
  title         = {Unlocking Understanding: An Investigation of Multimodal Communication in Virtual Reality Collaboration},
  booktitle     = {Proceedings of the CHI Conference on Human Factors in Computing Systems},
  series        = {CHI '24},
  year          = {2024},
  articleno     = {602},
  numpages      = {16},
  publisher     = {Association for Computing Machinery},
  address       = {New York, NY, USA},
  doi           = {10.1145/3613904.3642491},
}

@article{gloor2022entanglement,
  author        = {Gloor, Peter A. and Zylka, Matth{\"a}us P. and Fronzetti Colladon, Andrea and Makai, Marton},
  title         = {`Entanglement' -- A new dynamic metric to measure team flow},
  journal       = {Social Networks},
  volume        = {70},
  year          = {2022},
  pages         = {100--111},
  doi           = {10.1016/j.socnet.2021.11.010},
  url           = {https://doi.org/10.1016/j.socnet.2021.11.010},
}

@article{paulsen2024collaborativevr,
  author        = {Lucas Paulsen and Susanne Dau and Jacob Davidsen},
  title         = {Designing for collaborative learning in immersive virtual reality: a systematic literature review},
  journal       = {Virtual Reality},
  volume        = {28},
  number        = {1},
  year          = {2024},
  articleno     = {63},
  numpages      = {17},
  doi           = {10.1007/s10055-024-00975-4},
}

@article{delice2019advancing,
  author        = {Delice, Fabrice and Rousseau, Moira and Feitosa, Jennifer},
  title         = {Advancing teams research: What, when, and how to measure team dynamics over time},
  journal       = {Frontiers in Psychology},
  volume        = {10},
  year          = {2019},
  articleno     = {1324},
  numpages      = {20},
  publisher     = {Frontiers Media SA},
  doi           = {10.3389/fpsyg.2019.01324},
}

@article{lehmann2024multimodal,
  author        = {Lehmann-Willenbrock, Nale and Hung, Hayley},
  title         = {A multimodal social signal processing approach to team interactions},
  journal       = {Organizational Research Methods},
  volume        = {27},
  number        = {3},
  year          = {2024},
  pages         = {477--515},
  publisher     = {SAGE Publications},
  doi           = {10.1177/10944281231202741},
}

@article{enriquez2024layout,
  author        = {Enriquez, Daniel and Tong, Wai and North, Chris and Qu, Huamin and Yang, Yalong},
  title         = {Evaluating Layout Dimensionalities in {PC+VR} Asymmetric Collaborative Decision Making},
  journal       = {Proceedings of the ACM on Human-Computer Interaction},
  volume        = {8},
  number        = {ISS},
  year          = {2024},
  articleno     = {530},
  numpages      = {21},
  doi           = {10.1145/3698130},
}

@article{fink2024avatars,
  author        = {Fink, Daniel Immanuel and Skowronski, Moritz and Zagermann, Johannes and Reinschluessel, Anke Verena and Reiterer, Harald and Feuchtner, Tiare},
  title         = {There Is More to Avatars Than Visuals: Investigating Combinations of Visual and Auditory User Representations for Remote Collaboration in Augmented Reality},
  journal       = {Proceedings of the ACM on Human-Computer Interaction},
  volume        = {8},
  number        = {ISS},
  year          = {2024},
  articleno     = {548},
  numpages      = {29},
  doi           = {10.1145/3698148},
}

@article{van2018conceptualization,
  author        = {van den Hout, Jef J. J. and Davis, Orin C. and Weggeman, Mathieu C. D. P.},
  title         = {The conceptualization of team flow},
  journal       = {The Journal of Psychology},
  volume        = {152},
  number        = {6},
  year          = {2018},
  pages         = {388--423},
  publisher     = {Taylor \& Francis},
  doi           = {10.1080/00223980.2018.1449729},
  url           = {https://doi.org/10.1080/00223980.2018.1449729},
}

@misc{radford2022whisper,
  author        = {Alec Radford and Jong Wook Kim and Tao Xu and Greg Brockman and Christine McLeavey and Ilya Sutskever},
  title         = {Robust Speech Recognition via Large-Scale Weak Supervision},
  year          = {2022},
  doi           = {10.48550/arXiv.2212.04356},
  eprint        = {2212.04356},
  url           = {https://arxiv.org/abs/2212.04356},
  archivePrefix = {arXiv},
  primaryClass  = {eess.AS},
}

@misc{mohr2024bilingual,
  author        = {Isabelle Mohr and Markus Krimmel and Saba Sturua and Mohammad Kalim Akram and Andreas Koukounas and Michael G{\"u}nther and Georgios Mastrapas and Vinit Ravishankar and Joan Fontanals Mart{\'{i}}nez and Feng Wang and Qi Liu and Ziniu Yu and Jie Fu and Saahil Ognawala and Susana Guzman and Bo Wang and Maximilian Werk and Nan Wang and Han Xiao},
  title         = {Multi-Task Contrastive Learning for 8192-Token Bilingual Text Embeddings},
  year          = {2024},
  doi           = {10.48550/arXiv.2402.17016},
  eprint        = {2402.17016},
  url           = {https://arxiv.org/abs/2402.17016},
  archivePrefix = {arXiv},
  primaryClass  = {cs.CL},
}

@misc{gunther2024late,
  author        = {Michael G{\"u}nther and Isabelle Mohr and Daniel James Williams and Bo Wang and Han Xiao},
  title         = {Late Chunking: Contextual Chunk Embeddings Using Long-Context Embedding Models},
  year          = {2025},
  doi           = {10.48550/arXiv.2409.04701},
  eprint        = {2409.04701},
  url           = {https://arxiv.org/abs/2409.04701v3},
  note          = {Version 3, revised 7 July 2025},
  archivePrefix = {arXiv},
  primaryClass  = {cs.CL},
}

@misc{jinabaseDE2024,
  author        = {{Jina AI}},
  title         = {{jinaai/jina-embeddings-v2-base-de}},
  year          = {2024},
  howpublished  = {Hugging Face model card},
  url           = {https://huggingface.co/jinaai/jina-embeddings-v2-base-de},
  lastaccessed  = {October 17, 2026},
}

@misc{metaquestpro,
  author        = {{Meta}},
  title         = {{Meta Quest Pro}},
  year          = {[n.d.]},
  howpublished  = {Official product page},
  url           = {https://www.meta.com/quest/quest-pro/},
  lastaccessed  = {October 17, 2026},
}

@inproceedings{williamson2022digital,
  author        = {Williamson, Julie R. and O'Hagan, Joseph and Guerra-Gomez, John Alexis and Williamson, John H. and Cesar, Pablo and Shamma, David A.},
  title         = {Digital Proxemics: Designing Social and Collaborative Interaction in Virtual Environments},
  booktitle     = {CHI Conference on Human Factors in Computing Systems},
  series        = {CHI '22},
  year          = {2022},
  articleno     = {423},
  numpages      = {12},
  publisher     = {Association for Computing Machinery},
  address       = {New York, NY, USA},
  doi           = {10.1145/3491102.3517594},
  url           = {https://doi.org/10.1145/3491102.3517594},
}

@inproceedings{lechappe2025understanding,
  author        = {L{\'e}chapp{\'e}, Aur{\'e}lien and Johnstone, Ross and Milliat, Aur{\'e}lien and Williamson, John H. and Chollet, Mathieu and Williamson, Julie R.},
  title         = {Understanding Social Interactions in Reality Versus Virtuality},
  booktitle     = {Proceedings of the 2025 CHI Conference on Human Factors in Computing Systems},
  series        = {CHI '25},
  year          = {2025},
  articleno     = {41},
  numpages      = {18},
  publisher     = {Association for Computing Machinery},
  address       = {New York, NY, USA},
  doi           = {10.1145/3706598.3713628},
  url           = {https://doi.org/10.1145/3706598.3713628},
}

\end{document}